# A Real Time Optimistic Strategy to achieve Concurrency Control in Mobile Environments Using On-demand Multicasting

Salman Abdul Moiz[1] and Dr. Lakshmi Rajamani[2]

[1]Centre for Development of Advanced Computing, Bangalore, India
salman.abdul.moiz@ieee.org
[2]University College of Engineering, Osmania University, Hyderabad, India
drlakshmiraja@gmail.com

## ABSTRACT

In mobile database environments, multiple users may access similar data items irrespective of their physical location leading to concurrent access anomalies. As disconnections and mobility are the common characteristics in mobile environment, performing concurrent access to a particular data item leads to inconsistency. Most of the approaches use locking mechanisms to achieve concurrency control. However this leads to increase in blocking and abort rate. In this paper an optimistic concurrency control strategy using on-demand multicasting is proposed for mobile database environments which guarantees consistency and introduces application-specific conflict detection and resolution strategies. The simulation results specify increase in system throughput by reducing the transaction abort rates as compared to the other optimistic strategies proposed in literature.

## 1. INTRODUCTION

Mobile computing is widely used in many applications such as mobile banking, traffic status, weather forecasting, etc., [10, 5]. In order to provide these services, required information is retrieved from database server via a wireless channel and is passed on to the mobile hosts.

In literature two different approaches are used by which the fixed host (server) services the request of the mobile client's viz. On-demand method and broadcast method [1,3]. In the on-demand method, data are transmitted only when client's demand. This may lead to congestion and bottlenecks in upstream link as many mobile hosts may request different data items independently. In a wireless environment, generally, the downstream bandwidth is relatively high as compared to the upstream bandwidth [2]. In broadcast based methods, the fixed host transmits the data items to the mobile hosts periodically regardless of the client's demands.

In the broadcast-based method, the server transmits data items to the clients periodically regardless of their demands, then, the clients access and select the data items of interest through the broadcast channel [1, 2]. This method is known to be more efficient than the on-demand method in a mobile database environment. However this approach doesn't maintain data consistency as different mobile transactions may access the same data item concurrently through the broadcast channel [14]. Further the data which is of no importance is sent to few mobile hosts. Optimistic Concurrency control techniques detect and resolve data conflicts in the validation phase of the transaction execution. In most of the approaches the conflicting transactions are aborted in the validation phase.

In this paper, optimistic approach is presented where the data items are not locked and can be used by more than one mobile host at the same time. The transaction executes in two phases: In the first phase the transaction is committed locally on mobile host using the on-demand approach. In the second phase the results are updated onto the fixed host. Only those mobile hosts (multicasting) that were using the similar data items will be informed about the updated values by fixed host and the transaction will be restarted on the mobile host.

The remaining part of this paper is organized as follows: Section 2 summarizes the survey of optimistic concurrency control approaches, section 3 describes the environment and the elements of mobile databases, section 4 specifies the proposed concurrency control strategy, section 5 describes the behaviour of the proposed strategy and section 6 concludes the paper.

## 2. RELATED WORK

Concurrency Control is one of the important components of transaction management. Several valuable attempts were made to efficiently implement the concurrency control strategies in mobile environment.

Most of these proposals are based on three mechanisms viz., locking, timestamps and optimistic concurrency control. Though these schemes are well suited for traditional database applications, they don't work efficiently in mobile environments. Due to various constraints in the mobile environment and nature of different online applications, traditional concurrency control mechanism may not work effectively.

When two mobile clients access a data item concurrently where one client tries to read from the data item while the other tries to write upon it, it may result in inconsistency. For this purpose we might consider the two phase locking protocol, which requests the server to lock all the data items demanded. However, this protocol requires the clients to communicate continuously with the server to obtain the necessary locks and detect the data conflicts, and hence is not suitable to the wireless environment where the capacity of communication bandwidth is highly variable and unexpected disconnection may occur [6]. In order to avoid the problem of starvation due to locking, timeout based approach is proposed [15]. Every mobile client may not be able to execute the transaction within the specified time period, due to variation in bandwidth disconnection etc. Hence a dynamic timer adjustment strategy is proposed [16]. To reduce the frequent rollbacks in [15,16], a pre-emptive dynamic timer adjustment strategy[17] and a predictive strategy[18] is proposed. As the offline processing capability for individual mobile host may vary, it may execute the transaction faster even if it has requested for execution quiet later. For this reason, an optimistic concurrency control technique is frequently used in wireless environments [13, 14, 22].

Optimistic concurrency control protocols (OCC) [7, 9, 21] are non-blocking and deadlock-free, which make them efficient to use in mobile computing and have been adopted in the Disconnected Operation [11] and Kangaroo Transaction model [12]. However, without locks to data items, transactions might access conflicting data items under an optimistic concurrency control protocol (OCC). Two concurrent transactions conflict if one of them performs a write on similar data items. Therefore, approaches to terminate conflicting transactions are proposed [4, 6]. In these approaches if the conflict rate increases, more and more transactions get aborted. In [8], author proposes A Timestamp-Based Optimistic Concurrency Control for Handling Mobile Transactions". However it [8,20,22] needs broadcasting of messages to send the invalidation reports which might unnecessary flood the network thereby reducing the transaction throughput. The approach followed in [8] uses broadcasting which is not suitable for some of the applications because of the rise in abort rates and unnecessary flooding of invalidation reports. A Hybrid strategy to propagate the invalidation reports [19] based on AVI was proposed. However this strategy may still lead to blocking and suffers from computation overhead.

In this paper a mechanism to handle optimistic concurrency control is presented which doesn't terminate the transaction after detecting a conflict. Instead the new value of the conflicting data item is multicasted and the transaction is restarted with the new data items without the need to abort the transaction. Timestamp for the Global commit is maintained to know the time at which the new updated value is multicasted to conflicting transactions. This also save the uplink bandwidth as the mobile host may not request for the data items again for the execution of the transaction locally.

## 3. MOBILE DATABASE MODEL

### 3.1 Mobile Database Architecture

The mobile computing environment generally consists of three entities Fixed Host (FH), Mobile Hosts (MH) and Base Stations (BS) respectively. Terminals, desktop, servers are the fixed host, which are interconnected by means of a fixed network.

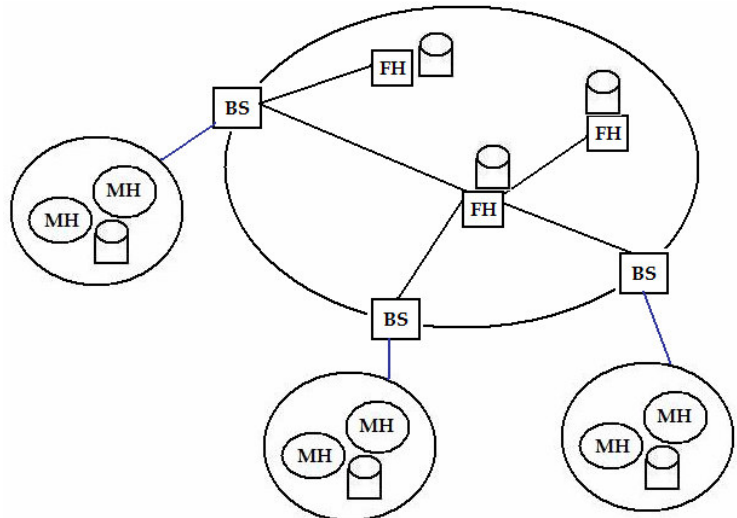

**Figure 1. Mobile Database Architecture**

Large databases can run on servers that guarantee efficient processing and reliable storage of database. Fixed hosts perform the transaction and data management functions with the help of data base servers (DBS). Mobile units are the portable computers which can retain the network connections through the support of the Base Stations (BS).

Transactions are initiated at a mobile host may be executed at fixed host or mobile host. A Mobile unit connects to a fixed host through a wireless link A Base station connects to a mobile unit and is equipped with a wireless interface. It is also known as a Mobile Support Station. Mobile Hosts (MH) may not always be connected to the fixed network. They may be disconnected for different reasons. Mobile host may differ with respect to the computing power and storage space; however MH can run a DBMS module.

In this paper we assume that the transaction is initiated at the mobile host. The data items needed to execute the transaction are copied into mobile host from the fixed host. The transaction is first executed locally. Once the transaction commits at mobile hosts the results are validated at fixed host. When the transaction is being executed locally, the mobile host may be disconnected and later the results may be reconciled with the fixed host.

### 3.2 Replication in Mobile Environment

Data Replication is the process that allows building a distributed environment through the management of multiple copies of data. Changes submitted to one replica have to be applied at the other replicas such that the different copies of the database remain consistent despite concurrent updates.

In general, there are two types of replication strategies: Synchronous and Asynchronous replication. In *Synchronous replication*, updates on a data item are performed on all replicas at the same time. In *Asynchronous replication*, write operations performed on one site is stored locally and later on it is updated to other replicas. Synchronous replication technology ensures highest level of data integrity but requires permanent availability of participating sites and transmission bandwidth. Asynchronous replication provides more flexibility than synchronous replication as single site could work even if a remote server is not reachable or down.

As disconnections are common characteristics in mobile environments, asynchronous replication is better suited. If the mobile host holding a replica is disconnected for a longer time then in synchronous replication the transaction can't proceed unless the mobile host is connected.

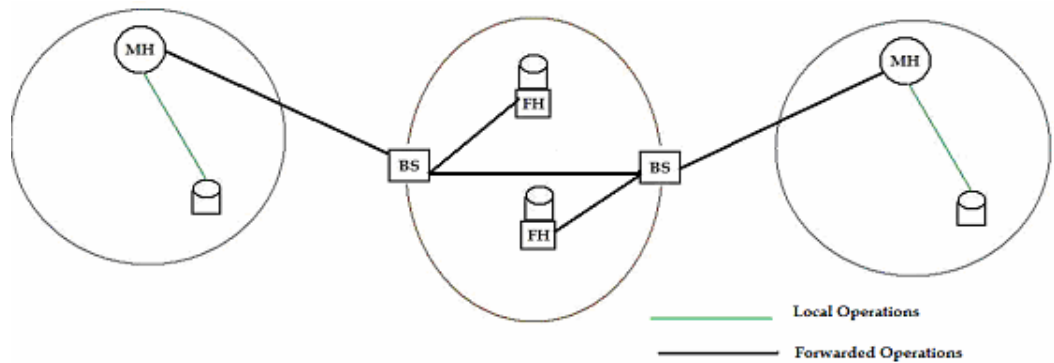

Figure 2. Working Scheme for Replication

In the proposed optimistic approach, a fragment is copied onto mobile host for execution of a transaction i.e. the transactions are first executed at mobile host and later the updates are propagated to fixed host. In figure 2, local operations are the operations which could be executed at mobile hosts and forwarded operations are the results propagated to fixed host. The decision for successful completion of a transaction is made only when the forwarded operations are successfully updated at fixed host. The fixed host acts as a master and the mobile hosts as slaves. Though the write operations are first performed locally by mobile host, but the transaction commits only when the write operation is successfully executed at fixed host. Hence Master-Slaves strategy is adopted in implementation of proposed concurrency control technique.

The fragment needed for execution of the transaction is copied at mobile host and the transaction is executed locally. This fragment is usually a mixed fragment, a combination of vertical and horizontal fragment. The locally committed transactions are forwarded to the fixed host to make a final commit decision.

When multiple mobile hosts request for the same data items, the respective data items are read from fixed host. When one mobile executes a transaction successfully and the results are updated at fixed host, Conflict may occur leading to concurrency violation.

Two operations conflict if they operate on same data item and one of the operations being a write operation (update conflict). As the mobile host has limited storage, the total relation is not replicated instead only the read/write operations are propagated between mobile host and fixed host. Hence the strategy used is transactional replication not the snapshot replication.

An efficient strategy is needed in case of asynchronous replication in order to detect and correct data conflicts due to concurrent modifications occurring at different mobile hosts between two database synchronization events.

## 4 OPTIMISTIC CONCURRENCY CONTROL STRATEGY

Concurrency control deals with the issues involved in allowing Simultaneous accesses to shared data items. Atomicity, consistency, and isolation of transactions are achieved in the database through concurrency control mechanisms. In particular, mobile applications have to face disconnections. It is expected that the transaction continues when the mobile host is disconnected. Hence there is a need of optimistic replication techniques.

In optimistic replication, shared data is replicated on mobile hosts and users are allowed to continue their work while disconnected. After successful completion of local operations at mobile host, the results are later propagated to fixed hosts. In the earlier approaches whenever a concurrency violation occurs i.e. data items are updated at fixed host the conflicting transaction using the similar data items was aborted. In this approach the conflicting transaction is not aborted but it is restated with new state of the data items.

### 4.1 Conflict Detection & Resolution

A transaction is initiated at mobile host and if it is committed the results are later on reconciled at the fixed host. The proposed strategy uses two phase. In tentative phase the transaction is executed at mobile host. If the transaction commits in tentative phase, then commit phase is initiated to perform the write operations on fixed host. During the Tentative Phase, the conflicts are detected when multiple hosts are accessing same data items. However in our approach the conflicts detected are tolerated till it enters into the Commit Phase.

The approach followed for conflict resolution is dynamic. This is because a mobile host which execute the transactions first will be committed. Hence the conflicts will be resolved when the write operations are propagated to the fixed host.

In a distributed environment, the commit protocol uses Coordinator and Participants for reaching to a final commit state. As mobile host is prone to disconnections and involves mobility it can't act as a coordinator. The Base Station (BS) to which a coordinator is registered can be used as coordinator which is responsible for Conflict resolution when the concurrency is violated.

Table 1. Elements of the Commit Protocol

| Participant | Coordinator | Participant |
|---|---|---|
| Mobile Host | Base Station | DBS |

Since the Commit decision is made by the coordinator, even if the mobile host moves from one cell to another cell consistency is preserved. If mobile host moves from cell 1 to cell 2, the information regarding the Base station to which the mobile was registered is communicated to the base station of next cell to which the mobile host is registered. If the transaction is executed successfully in cell 2, the result of the sub transactions is sent to the cell1 base station to make the final commit decision. To make a final commit decision when the mobile is moving, inter base station communication is needed to reach the final state.

### 4.1.1   Tentative Phase

- Mobile host request for a particular data item for executing a particular transaction identified by TransactionId or name (1)
- The fixed host checks for the required data items needed to execute a transaction (2)

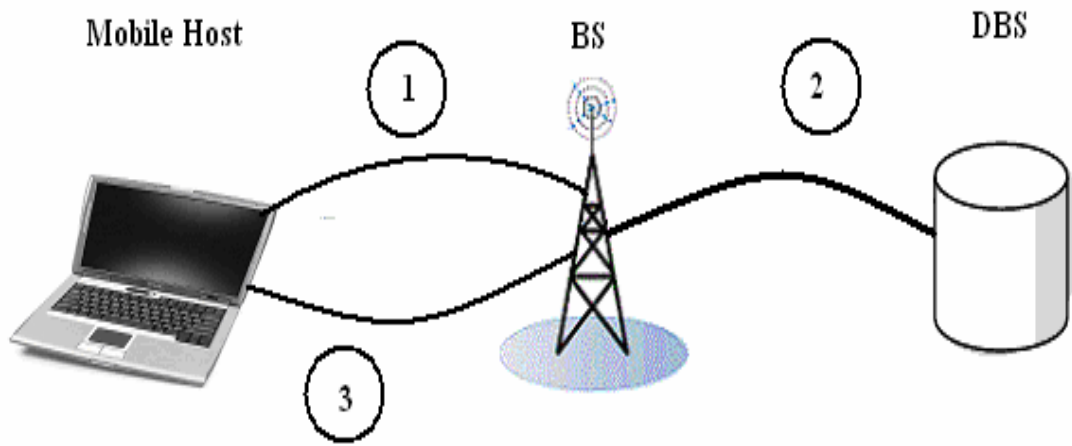

**Figure 3. Communication among mobile host and fixed host in tentative phase**

- The information of current transaction is recorded by the base station in current transactions relations. The state of shared data item in particular will help in identifying the conflicts.
- The base station also scans the current transactions relations to know whether any other mobile host has already requested for the same data items. This helps in detecting the conflicts. However this is tolerated till the commit or validation phase. If so the Arrival_Time of the transaction which arrived first is multicasted to mobile host which requested for the same data items.
- The data items needed to execute the transaction are read by the mobile host (3)
- If the transaction commits got commit phase

### 4.1.2   Commit Phase

- Once transaction commits locally, the results are propagated to the coordinator(BS) for making the final commit decision.(4)

- The coordinator scans the Current transactions relations to check whether any other mobile host is using the similar data items which are updated by committed transactions.
- If so the updated values are propagated to the mobile hosts by updating the Arrival_Time and requesting to restart the transaction. (7). (Conflict Resolution).
- The coordinator Base station updates the data items and removes that entry from Current transactions relations (5) and informs the mobile host about the final commit decision (4)

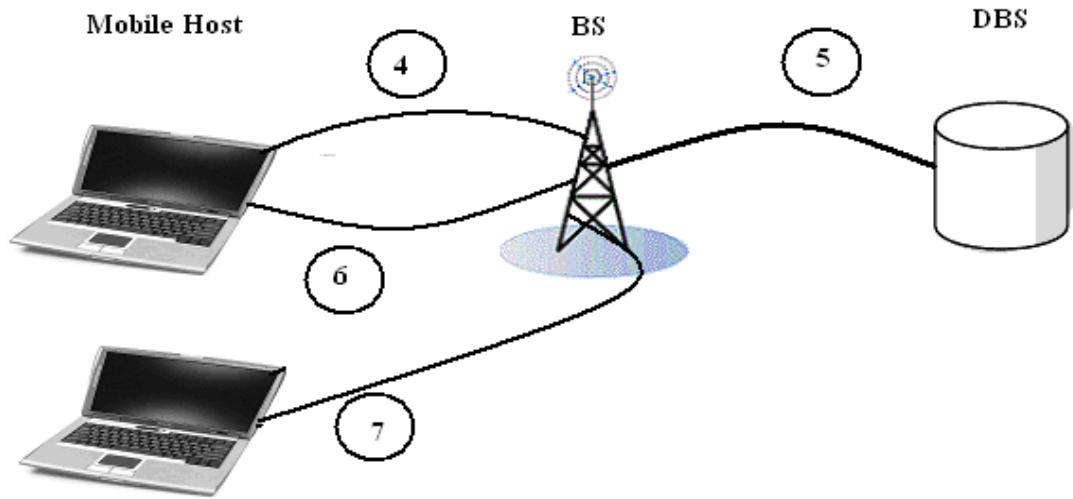

**Figure 4. Communication among Fixed host and Mobile hosts in commit phase**

There are basically two ways to perform the validation for detecting data conflicts, *backward validation* and *forward validation*. Backward validation examines data items written by recently committed transactions with data items of the transaction in question, whereas forward validation resolves conflicts by examining data items of the transaction in question with those data items read by the transactions currently active in the *write* phase [8]. The backward validation technique seems is appropriate because it tries to resolve conflicts among already committed transactions.

Though the mobile host which completed the tentative phase successfully can directly send the updated values to another site executing the same transaction. However in order to avoid multiple updates, we use the single control at fixed host. Further the fixed host has a mirror copy such that in case of its failure the backup copy will be active.

### 4.2 Data Structures (Fixed Host)

In order to avoid blocking and to enhance the throughput for optimistic concurrency control, it is proposed that the fixed host maintains 2 relations in addition to the base table

### 4.1.1 Transaction Info relation

This relation stores the list of possible transactions which can be executed on mobile host

**Table 2. Structure of Transaction_Info relation**

| TransactionId | Name | Relation | Data Item(s) |
|---|---|---|---|

*TransactionId*   : Every transaction has a unique Id

*Name*           : A transaction can also be identified by name. For example, for a mobile banking the name of the transactions could be transfer amt, withdraw amt etc.

*Relation*        : This gives the name of the table whose data items are needed to execute a transaction identified by TransactionId or Name

*DataItem(s)*     : This specifies list of data items needed from the relation to execute the transaction identified by TransactionId.

The advantage of using this relation is to reduce the lookup operation needed to know the data items for execution of a transaction. Secondly this helps in quick extraction of the mixed fragment needed to execute a transaction at mobile host.

### 4.1.2 Current Transactions relation.

This table contains the list of current transactions executed at site i

**Table 3. Structure of Current Transactions Relation**

| Site_Id | TransactionId | Data_Items | Arrival Time |
|---|---|---|---|

*Site id*         :  represents a particular mobile host i which requested for execution of Transaction identified by Transaction_id
*TransactionId*   : represents type of transaction

*Data_Items*      :  represents the shared items, whose values may conflict.

*Arrival Time*    : Specifies the time at which the data items are copied into mobile host for execution.

This relation is used keep track of all sites, time at which data items were copied at mobile host. This helps the coordinator in sending the modified state of the data items as a result of successful completion of a transaction at a mobile host $m_i$. It is also helpful for mobile hosts in knowing the arrival time of the conflicting transactions so that the mobile hosts which acquired the data items later has the knowledge that it may have to re-execute the transaction if new updates are propagated.

The advantage of this approach is due to variable bandwidth and disconnections in mobile environments, there might be a possibility that the data items which were acquired by a particular mobile host quiet later, may commit the transaction, prior to the mobile host which has requested the data item first. In this case the coordinator sends the new values of the data items to mobile host which is not yet committed and updates the Arrival Time.

## 5  PERFORMANCE METRICS

The proposed optimistic concurrency control is simulated using, postgress as the database. The front-end differs from application to application. A front-end for mobile banking with basic transactions is designed. The simulator follows MVC (Model-View- Controller) architecture. The functionality of the coordinator i.e the Base station is implemented using J2EE at fixed host. The system is finally tested using 10 different types of transactions involving concurrent requests for execution of a transaction.

In a mobile banking application few of the transactions can be executed by sending a request from the mobile host. Consider the following relation which contains the account holder's information.

**Table 4. Account holders Information**

| Account_no | Amount |
|---|---|
| 101 | 10000 |
| 102 | 12300 |
| 103 | 11500 |

The base station is maintained at Data base server which is one of the participant of the commit protocol. The coordinator i.e base station maintains two relations which helps in detecting the conflicts and resolving the same.

**Table 5. Transaction Info relation**

| TransactionId | Name | Relation | Data Item(s) |
|---|---|---|---|
| T1 | Deposit | Account | Account_no, Amount |
| T2 | Withdraw | Account | Account_no, Amount |
| T3 | Enquiry | Account | Amount |

Table 5 provides a list of possible transactions i.e Deposit, Withdraw and Enquiry, the base relation to be used data items needed for the execution

Assume that two joint account holders with Account number 103 issued a request for deposit (Rs.1000/- ) and withdrawal (Rs. 500/-) of amount from mobile host M1 and M2 respectively with a difference of 5 sec (say). They also provide their account number to retrieve a particular tuple.

**Table 6. Current transaction entries when M1 gave a request**

| Site_Id | TransactionId | Account_no | Amount | Arrival Time |
|---------|---------------|------------|--------|--------------|
| M1      | T1            | 103        | 11500  | 10.05        |

When M1 requested for execution of a transaction with transactionId T1, the coordinator sends Account_no and Amount to M1. The same information is entered in current transaction relation. The transact T1 starts executing at M1.

M2 now request for depositing amount after 5 secs with transaction id and account number. Depending on the transaction to be performed using Transaction id, the respective data items are copied into current transaction. In the current transaction relation M1 has also requested for a transaction on same account no. M2 is now aware that M1 has requested for the same data item 5 mins back

**Table 7. Current transaction entries after M2 initiated the request**

| Site_Id | TransactionId | Account_no | Amount | Arrival Time |
|---------|---------------|------------|--------|--------------|
| M1      | T1            | 103        | 11500  | 10.05        |
| M2      | T2            | 103        | 11500  | 10.10        |

Both M1 and M2 are now executing the transaction with the data item replicated at their respective mobile host. Once it completes the tentative phase successfully, it enters into commit phase.

Case (i): M1 completes tentative phase before M2

When M1 completes the execution before M2, the final value of M1 after deposit operation i.e Rs. 12500/- will be updated in base table. At the same time the updated values is propagated to M2 to restart the transaction with new value at the same time arrival time is set to current time. Further the entry of M1 is deleted from Current transaction relation. The transaction initiated by M2 is not aborted in commit phase. But when concurrency violation occurs, the new value for the data item is propagated only to those mobile hosts using the shared data item. This is unlike most of the optimistic approaches where the data is broadcasted after finite period of time.

Case(ii) M2 Completes tentative phase before M1

In mobile environments, due to variable bandwidths and disconnections, there is a possibility that the transaction which started later can commit first. In this scenario if M2 completes the execution first, the final value 11000/- is updated in base table at DBS. As conflict is detected, the new value is propagated to M1. The transaction restarts at M1 and T2 is discarded from the current transaction relation.

Case (iii) M1 and M2 completes at the same time

There is a possibility that both M1 and M2 may complete the transaction at the same time and enter into a commit phase with an updated value. In this scenario the transaction which requested for the data item first will be committed and the other mobile host is required to re-execute the transaction with new value for the shared data item.

In all the scenarios, Serializability is preserved because the final commit decision is made by the coordinator and the data is updated in DBS based on certain serial order of execution of transactions.

## 6  CONCLUSION

The Optimistic Concurrency Control Strategy doesn't use any locking which doesn't block the shared resources. Further concurrency can be guaranteed by first executing transactions locally and later on propagating the results. In this scheme whenever a fixed host detects a concurrency violation, it propagates the updated shared data item to the mobile host using the same data item without aborting it. The mobile host which successfully completes the transaction locally will be committed irrespective of its arrival time. In this scheme there could be a possibility that the transaction which arrived quiet early might not get executed because the other mobile hosts are executing faster.  The future work may introduce a priority field to give chance to the transaction which requested first or a hybrid approach for concurrency control that enters uses pessimistic strategy by partially locking few data items to complete its execution.

**Authors**

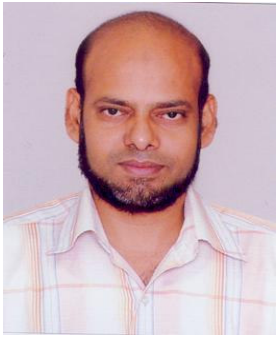

**Salman Abdul Moiz** is a Research Scientist at Centre for Development of Advanced Computing, Bangalore. He received his B.Sc from Osmania University, MCA from Osmania University, M.Tech(cse) from Osmania University and M.Phil(Cs) from Madurai Kamaraj University.

He is a Research Scholar at Osmania University and published more than 20 papers in various National/International Conferences and Journals. His areas of interests include Mobile databases, Software Process Improvements, Agile Methodology & Disaster Recovery.

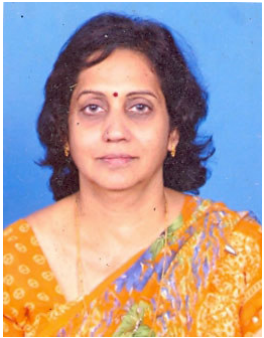

**Dr. Lakshmi Rajamani** is working as Professor & Head of the Department, CSE, University College of Engineering, Osmania University, Hyderabad. She received M.Sc (Statistics) from IIT Kanpur, M.Phil (Computer methods) from University of Hyderabad and PhD (CSE) from Jadavpur University, Kolkata. She authored more than 25 papers in various National/International conferences and Journals. Her research interests are in the areas of Neural Networks, Artificial Intelligence, Distributed Computing & Data Mining. She worked as Chairperson, Board of Studies, CSE, Osmania University. She also worked as Director CDAC, University College of Engineering, Osmania University.